\documentclass[preprint]{elsarticle}
\usepackage{lineno,hyperref,natbib}

\usepackage{amssymb}

\journal{Nuclear Physics A}

\begin{document}

\begin{frontmatter}



\title{Particle-number conserving analysis of the high-spin structure of $^{159}$Ho}



\author{Zhen-Hua Zhang\fnref{contact}}
\ead{zhzhang@ncepu.edu.cn}
\address{Mathematics and Physics Department,
              North China Electric Power University, Beijing 102206, China}

\begin{abstract}
The high-spin rotational bands in odd-$Z$ nuclei $^{159}$Ho ($Z=67$) are investigated
using the cranked shell model with the pairing correlations treated
by a particle-number conserving method,
in which the blocking effects are taken into account exactly.
The experimental moments of inertia and alignments
and their variations with the rotational frequency $\hbar\omega$
are reproduced very well by the calculations.
The splitting between the signature partners of the
yrast band $7/2^-[523]$ is discussed and the splitting of
the excited band $7/2^+[404]$ above $\hbar\omega \sim 0.30$~MeV
is predicted due to the level crossing with $1/2^+[411]$.
The calculated $B(E2)$ transition probabilities
are also suggested for future experiments.
\end{abstract}


\end{frontmatter}


\section{Introduction}
\label{introduction}

The investigations of the nuclear structure toward the extremes of
angular momentum in the rare-earth nuclei have resulted in the
discovery of various new phenomena, e.g.,
back-bending~\cite{Johnson1971_PLB34-605},
band termination~\cite{Bengtsson1983_PST5-165},
signature inversion~\cite{Bengtsson1984_NPA415-189},
super-deformed band~\cite{Twin1986_PRL57-811}
wobbling motions in the triaxial strongly deformed nuclei~\cite{Odegard2001_PRL86-5866}
et al., which make this mass region an excellent testing ground
for various nuclear models concerning on the fast rotating nucleus, e.g.,
the cranked Nilsson-Strutinsky method~\cite{Andersson1976_NPA268-205},
the Hartree-Fock-Bogoliubov cranking model with Nilsson~\cite{Bengtsson1979_NPA327-139}
and Woods-Saxon potentials~\cite{Nazarewicz1985_NPA435-397, Cwiok1987_CPC46-379},
the projected shell model~\cite{Hara1995_IJMPE4-637},
the tilted axis cranking model~\cite{Frauendorf2001_RMP73-463},
the cranked relativistic~\cite{Afanasjev1996_NPA608-107}
and non-relativistic mean-field models~\cite{Dobaczewski1997_CPC102-166} et al.
The transitional rare-earth nuclei with $N \sim 90$ and $A \sim 160$
are particularly rich in such phenomena.
In this mass region, the nuclei are prolate deformed and
with increasing angular momentum, the yrast structure contains
a great number of rotational bands based on different
quasiparticle configurations.
Therefore, a complex picture of band crossings emerges due to rotation.
One pioneer work is the backbending first discovered
in $^{160}$Dy~\cite{Johnson1971_PLB34-605}.
Later on, it was interpreted as an alignment of one pair of
$i_{13/2}$ neutrons~\cite{Stephens1972_NPA183-257}.
Determining the nature of the backbending can help identify the
quasiparticle configurations along the yrast line.

Previous investigations show that the mean-field models are capable of explaining
the effects such as presence and absence
of the backbending in the neighboring nuclei~\cite{Dudek1988_PRC38-940}
or double backbending due to the coexistence of the neutron
and proton alignments~\cite{Cwiok1980_NPA333-139}.
For a review, please see Ref.~\cite{Voigt1983_RMP55-949}.
Usually, in most of these models, pairing correlations are treated by
Bardeen-Cooper-Schrieffer (BCS) or Hartree-Fock-Bogolyubov (HFB) formalism.
Now the BCS or HFB approximations have become standard methods in nuclear physics.
However, along with their great success,
both of them raise some concerns~\cite{Zeng1983_NPA405-1, Molique1997_PRC56-1795},
non-conservation of the particle-number being one of them.
Actually, all cranked HFB calculations show that a pairing collapsing occurs for angular
momentum $I$ greater than a critical value $I_c$.
In other words, the paring interaction has no effect in the high-spin region
at $I > I_c$ in these models.
The remedy in terms of the particle-number projection or the
Lipkin-Nogami method can restore this broken symmetry.
Previous investigations show that, after performing the particle-number projection,
the description of the rotational properties can be improved considerably
comparing with the HFB cranking calculations~\cite{Dudek1988_PRC38-940}.
However, they complicate the algorithms considerably, yet without improving the
description of the higher-excited part
of the spectrum of the pairing Hamiltonian~\cite{Molique1997_PRC56-1795}.
It is well known that pairing correlations are extremely
important in the low angular momentum region, where they are
manifested by reducing the nuclear moment of inertia (MOI) of the
rigid-body estimation.
Further investigation indicates that, at the high-spin region~($\hbar\omega \sim 0.8$~MeV),
although the MOI of rotational bands tends to be the same with or without pairing interaction,
the backbending frequencies still have large differences~\cite{Liu2009_PRC80-044329}.
Therefore, even at the high-spin region,
the pairing correlations should be included and treated correctly in the
theoretical models in order to understand some nuclear phenomena correctly.

As a typical example, the observed low-lying 1-quasiparticle bands in
$^{159}$Ho~\cite{Ma2000_JPG26-43, Ollier2011_PRC84-027302}
are analyzed by the cranked shell model with the pairing correlations treated
by a particle-number conserving method~\cite{Zeng1994_PRC50-1388}.
In contrary to the conventional BCS or HFB approach,
in the particle-number conserving method, the cranked shell model Hamiltonian
is solved directly in a truncated Fock-space~\cite{Wu1989_PRC39-666}.
Therefore the particle-number is conserved and the Pauli blocking effects
are taken into account exactly.
The particle-number conserving method has been used to describe successfully the normally
deformed and superdeformed high-spin rotational bands of nuclei
with $A \approx$ 160, 190, and
250~\cite{Zeng1994_PRC50-746, Liu2002_PRC66-024320, He2005_NPA760-263,
Zhang2009_NPA816-19, Zhang2009_PRC80-034313, He2009_NPA817-45,
Zhang2011_PRC83-011304R, Zhang2012_PRC85-014324, Zhang2013_PRC87-054308}.
The particle-number conserving scheme has also been implemented both in relativistic
and nonrelativistic mean field models~\cite{Meng2006_FPC1-38, Pillet2002_NPA697-141}
in which the single-particle states are calculated from self-consistent
mean field potentials instead of the Nilsson potential.
Recently, the particle-number conserving methods based on the
total-Routhian-surface method with the Woods-Saxon
potential~\cite{Fu2013_PRC87-044319} and
the cranking Skyrme-Hartree-Fock model have
been developed~\cite{Liang2015_PRC92-064325}.

The paper is organized as follows.
A brief introduction of the particle-number conserving method
for the cranked shell model is presented in Sec.~\ref{sec:PNC-CSM}.
The numerical details are given in Sec.~\ref{sec:num}.
The results and discussion are given in Sec.~\ref{sec:result}.
Finally, a summary is given in Sec.~\ref{sec:summ}.

\section{A brief introduction to particle-number conserving method for the cranked shell model}
\label{sec:PNC-CSM}

The cranked shell model Hamiltonian of an axially symmetric nucleus in the rotating
frame can be written as
\begin{eqnarray}
 H_\mathrm{CSM}
 & = &
 H_0 + H_\mathrm{P}
 = H_{\rm Nil}-\omega J_x + H_\mathrm{P}
 \ ,
 \label{eq:H_CSM}
\end{eqnarray}
where $H_{\rm Nil}$ is the Nilsson Hamiltonian, $-\omega J_x$ is the
Coriolis interaction with cranking frequency $\omega$ about the $x$
axis, perpendicular to the nuclear symmetrical $z$ axis, $H_0=H_{\rm
Nil}-\omega J_x$ is the one-body part of $H_{\rm CSM}$, and
$H_{\rm P} = H_{\rm P}(0) + H_{\rm P}(2)$ is the pairing interaction
\begin{eqnarray}
 H_{\rm P}(0)
 & = &
  -G_{0} \sum_{\xi\eta} a^\dag_{\xi} a^\dag_{\bar{\xi}}
                        a_{\bar{\eta}} a_{\eta}
  \ ,
 \\
 H_{\rm P}(2)
 & = &
  -G_{2} \sum_{\xi\eta} q_{2}(\xi)q_{2}(\eta)
                        a^\dag_{\xi} a^\dag_{\bar{\xi}}
                        a_{\bar{\eta}} a_{\eta}
  \ ,
\end{eqnarray}
where $\bar{\xi}$ ($\bar{\eta}$) labels the time-reversed state of a
Nilsson state $\xi$ ($\eta$), $q_{2}(\xi) = \sqrt{{16\pi}/{5}}
\langle \xi |r^{2}Y_{20} | \xi \rangle$ is the diagonal element of
the stretched quadrupole operator, and $G_0$ and $G_2$ are the
effective strengths of monopole and quadrupole pairing interactions, respectively.

Instead of the usual single-particle level truncation in conventional
shell-model calculations, a cranked many-particle configuration
truncation (Fock space truncation) is adopted which is crucial
to make the particle-number conserving calculations for low-lying excited states both
workable and sufficiently accurate~\cite{Molique1997_PRC56-1795, Wu1989_PRC39-666}.
Usually a dimension of 1000 should be enough for the calculations of heavy nuclei.
Previous investigation shows that when similar computational work is spent,
the results obtained in the many-particle configuration truncation
are much more accurate and reliable than those obtained
in the single-particle level truncation~\cite{Wu1989_PRC39-666}.
An eigenstate of $H_\mathrm{CSM}$ can be written as
\begin{equation}
 |\Psi\rangle = \sum_{i} C_i \left| i \right\rangle
 \qquad (C_i \; \textrm{real}) \ ,
\end{equation}
where $ |i\rangle $ is an eigenstate of $H_0$, i.e., a cranked many-particle configuration.
By diagonalizing the $H_\mathrm{CSM}$ in a sufficiently
large cranked many-particle configuration space,
sufficiently accurate solutions for low-lying excited eigenstates of
$H_\mathrm{CSM}$ are obtained.

The angular momentum alignment for the state $| \Psi \rangle$ is
\begin{equation}
\langle \Psi | J_x | \Psi \rangle = \sum_i C_i^2 \langle i | J_x | i
\rangle + 2\sum_{i<j}C_i C_j \langle i | J_x | j \rangle \ ,
\end{equation}
and the kinematic moment of inertia of state $| \psi \rangle$ is
\begin{equation}
J^{(1)}=\frac{1}{\omega} \langle\Psi | J_x | \Psi \rangle \ .
\end{equation}
Because $J_x$ is a one-body operator, $\langle i | J_x | j \rangle$
($i\neq j$) may not vanish when two cranked many-particle configurations $|i\rangle$ and
$|j\rangle$ differ by only one particle occupation. After a certain
permutation of creation operators, $|i\rangle$ and $|j\rangle$ can
be recast into
\begin{equation}
 |i\rangle=(-1)^{M_{i\mu}}|\mu\cdots \rangle \ , \qquad
|j\rangle=(-1)^{M_{j\nu}}|\nu\cdots \rangle \ ,
\end{equation}
where $\mu$ and $\nu$ denotes two different single-particle states,
the ellipsis $\cdots$ stands for the same particle occupation,
and $(-1)^{M_{i\mu}}=\pm1$, $(-1)^{M_{j\nu}}=\pm1$ according to
whether the permutation is even or odd.
Therefore, the angular momentum alignment of
$|\Psi\rangle$ can be expressed as
\begin{equation}
 \langle \Psi | J_x | \Psi \rangle = \sum_{\mu} j_x(\mu) + \sum_{\mu<\nu} j_x(\mu\nu)
 \ .
 \label{eq:jx}
\end{equation}
where the diagonal contribution $j_x(\mu)$ and the
off-diagonal (interference) contribution $j_x(\mu\nu)$ can be written as
\begin{eqnarray}
j_x(\mu)&=&\langle\mu|j_{x}|\mu\rangle n_{\mu} \ ,
\\
j_x(\mu\nu)&=&2\langle\mu|j_{x}|\nu\rangle\sum_{i<j}(-1)^{M_{i\mu}+M_{j\nu}}C_{i}C_{j}
  \qquad  (\mu\neq\nu) \ ,
\end{eqnarray}
and
\begin{equation}
n_{\mu}=\sum_{i}|C_{i}|^{2}P_{i\mu} \ ,
\end{equation}
is the occupation probability of the cranked orbital $|\mu\rangle$,
$P_{i\mu}=1$ if $|\mu\rangle$ is occupied in $|i\rangle$, and
$P_{i\mu}=0$ otherwise.

The $B(E2)$ transition probabilities can be derived
in the semiclassical approximation as~\cite{Bohr1975_Book}
\begin{equation}
B(E2) = \frac{3}{8}
{\left\langle \Psi | Q_{20}^{\rm p} | \Psi \right\rangle}^2 \ ,
 \label{eq:be2}
\end{equation}
where $Q_{20}^{\rm p}$ correspond to
the quadrupole moments of protons and
\begin{equation}
Q_{20} = \sqrt{\frac{5}{16\pi}} (3z^2-r^2) = r^2 Y_{20} \ .
\end{equation}

The experimental kinematic moment of inertia for each band is extracted by
\begin{equation}
\frac{J^{(1)}(I)}{\hbar^2}=\frac{2I+1}{E_{\gamma}(I+1\rightarrow
I-1)} \
\end{equation}
separately for each signature sequence within a rotational band
($\alpha$= I mod 2). The relation between the rotational frequency
$\omega$ and nuclear angular momentum $I$ is
\begin{equation}
\hbar\omega(I)=\frac{E_{\gamma}(I+1\rightarrow
I-1)}{I_{x}(I+1)-I_{x}(I-1)} \ ,
\end{equation}
where $I_{x}(I)=\sqrt{(I+1/2)^{2}-K^{2}}$, $K$ is the projection of
nuclear total angular momentum along the symmetry $z$ axis of an
axially symmetric nuclei.

\section{Numerical details}
\label{sec:num}

In this work, the deformation parameters $\varepsilon_2= 0.227$ and $\varepsilon_4=-0.0205$
are taken from Ref.~\cite{Bengtsson1986_ADNDT35-15},
which are taken as an average of the neighboring even-even
Dy ($Z=66$) and Er ($Z=68$) isotopes.
The Nilsson parameters are taken from the Lund systematics
($\kappa_{\rm p}$ = 0.0642, $\mu_{\rm p} = 0.596$ for protons and
$\kappa_{\rm n}$ = 0.0637, $\mu_{\rm n} = 0.428$ for neutrons)~\cite{Bengtsson1985_NPA436-14}
and a slight change [for proton $\mu_5=0.546$ ($N=5$ major shell) and for neutrons $\mu_5=0.340$]
is made to account for the experimental bandhead energies of
1-quasiparticle bands (see Tab.~\ref{tab:lev}).
In addition, the proton orbital $1/2^{-}[541]$ is shifted upward by 0.050$\hbar\omega_{0}$.

The effective pairing strengths, in principle, can be determined by
the odd-even differences in binding energies, and are connected with
the dimension of the truncated cranked many-particle configuration space.
In this work, the cranked many-particle configuration space is constructed in the proton $N=0-5$ shells
and the neutron $N=0-6$ shells.
The cranked many-particle configuration truncation energies are
about 0.8$\hbar\omega_0$ for protons and 0.7$\hbar\omega_0$ for neutrons.
For $^{159}$Ho, $\hbar\omega_{\rm 0p}=7.171$~MeV
for protons and $\hbar\omega_{\rm 0n}=7.965$~MeV for neutrons~\cite{Nilsson1969_NPA131-1}.
The dimensions of the cranked many-particle configuration space are
about 1500 for both protons and neutrons in the calculation.
The corresponding effective monopole and quadrupole pairing strengths are
$G_{\rm 0p}=0.47$~MeV and $G_{\rm 2p}=0.010$~MeV for protons,
$G_{\rm 0n}=0.32$~MeV and $G_{\rm 2n}=0.008$~MeV for neutrons.
Previous investigations have shown that
after the quadrupole pairing included,
the description of experimental band-head energies
and the level crossing frequencies can be improved~\cite{Diebel1984_NPA419-221}.

The stability of the calculations against the change of the dimension of
the cranked many-particle configuration space has been investigated in
Refs.~\cite{Molique1997_PRC56-1795,Zeng1994_PRC50-1388}.
In the present calculations, almost all the
cranked many-particle configurations with weight $>0.1\%$ are taken into account, so the solutions
to the low-lying excited states are accurate enough.
A larger cranked many-particle configuration space with renormalized
pairing strengths gives essentially the same results.

\section{Results and discussion}
\label{sec:result}

It is well known that the traditional parameters
($\kappa$ and $\mu$)~\cite{Nilsson1969_NPA131-1} used in the Nilsson Hamiltonian
have been quite successful in describing the single-particle structure
for stable nuclei, especially for those in the rare-earth and the actinide region.
However, for the specific nucleus, this level scheme is unable
to correctly reproduce the experimental bandhead energies of the
low-lying excited 1-quasiparticle bands, which is very important for investigating the
rotational properties of these nuclei, e.g., the level crossing frequencies et al.
To illustrated this more obviously, the experimental and the calculated bandhead
energies of 1-quasiparticle bands in $^{159}$Ho are shown in Tab.~\ref{tab:lev}.
The calculated results using traditional
Nilsson parameters~\cite{Nilsson1969_NPA131-1}
and the modified values are denoted by $E_{\rm Cal}^{\rm Lund}$ and
$E_{\rm Cal}^{\rm New}$, respectively.
It can be seen that the calculated 1-quasiparticle excitation energies
using Lund systematics are much higher than the experimental values.
After slightly adjusting the Nilsson parameters (see Sec.~\ref{sec:num}),
all the calculated results are lowered about 200~keV
and are more consistent with the data except $5/2^+[402]$, whose excitation energy
is still higher than the experimental value.
If the Nilsson parameters ($\kappa$ and $\mu$) were adjusted more sophisticated,
the calculated results may be improved further.
In the following calculation, this slightly modified Nilsson level scheme will be adopted.

\begin{figure}[h]
\centering
\includegraphics[width=0.8\columnwidth]{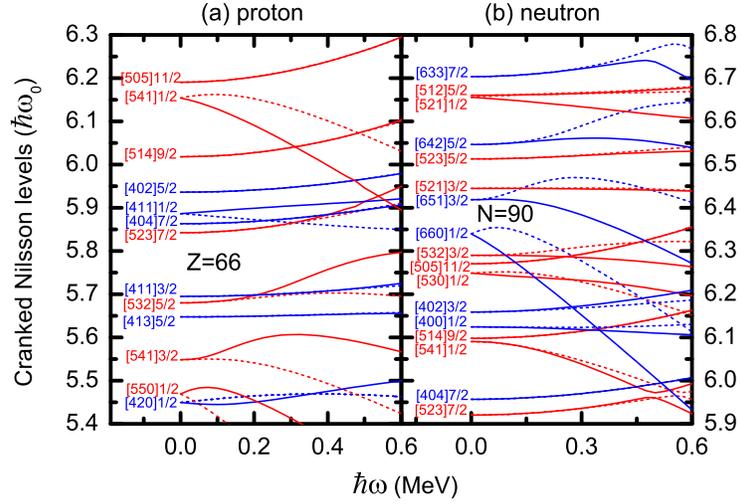}
\caption{\label{fig1:Nilsson} (Color online)
The cranked Nilsson levels near the Fermi surface of $^{159}$Ho
(a) for protons and (b) for neutrons.
The positive (negative) parity levels are denoted by blue (red) lines.
The signature $\alpha=+1/2$ ($\alpha=-1/2$) levels are denoted by solid (dotted) lines.
The deformation parameters $\varepsilon_2= 0.227$ and $\varepsilon_4=-0.0205$
are taken from Ref.~\cite{Bengtsson1986_ADNDT35-15},
which are taken as an average of the neighboring even-even Dy and Er isotopes.
The Nilsson parameters are taken from the Lund systematics values
($\kappa_{\rm p}$ = 0.0642, $\mu_{\rm p} = 0.596$ for protons and
$\kappa_{\rm n}$ = 0.0637, $\mu_{\rm n} = 0.428$ for neutrons)~\cite{Nilsson1969_NPA131-1}
and a slight change [for proton $\mu_5=0.546$ ($N=5$ major shell) and for neutrons $\mu_5=0.340$]
is made to account for the experimental bandhead energies of 1-quasiparticle bands.
In addition, the proton orbital $1/2^{-}[541]$ is shifted upward by 0.050$\hbar\omega_{0}$.
}
\end{figure}

In Fig.~\ref{fig1:Nilsson}, the cranked Nilsson levels near the Fermi surface of $^{159}$Ho
(a) for protons and (b) for neutrons are shown.
The positive (negative) parity levels are denoted by blue (red) lines.
The signature $\alpha=+1/2$ ($\alpha=-1/2$) levels are denoted by solid (dotted) lines.
Fig.~\ref{fig1:Nilsson} shows that near the Fermi surface there exists a proton sub-shell at $Z=66$
while the neutron sub-shell structure is not very clear.

\begin{figure}[h]
\centering
\includegraphics[width=0.8\columnwidth]{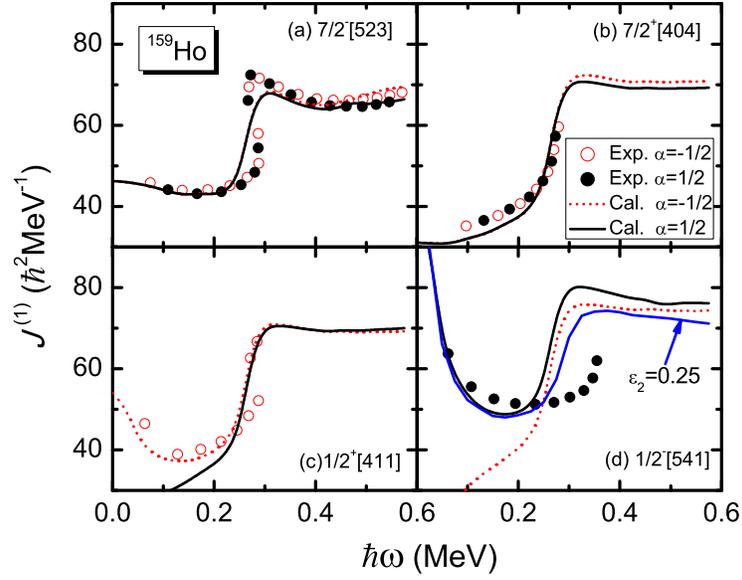}
\caption{\label{fig2:MOI} (Color online)
The experimental and calculated kinematic MOI's $J^{(1)}$ of four low-lying
bands in $^{159}$Ho.
The experimental MOI's, which are taken from Refs.~\cite{Ma2000_JPG26-43, Ollier2011_PRC84-027302},
are denoted by black solid circles (signature $\alpha=+1/2)$
and red open circles (signature $\alpha=-1/2)$, respectively.
The calculated MOI's by the particle-number conserving method are denoted by
black solid lines (signature $\alpha=+1/2)$
and red dotted lines (signature $\alpha=-1/2)$, respectively.
The calculated MOI's for  $1/2^-[541]$ ($\alpha =+1/2$) with the deformation
$\varepsilon_2 = 0.25$ are denoted by blue solid line.
}
\end{figure}

\begin{figure}[h]
\centering
\includegraphics[width=0.8\columnwidth]{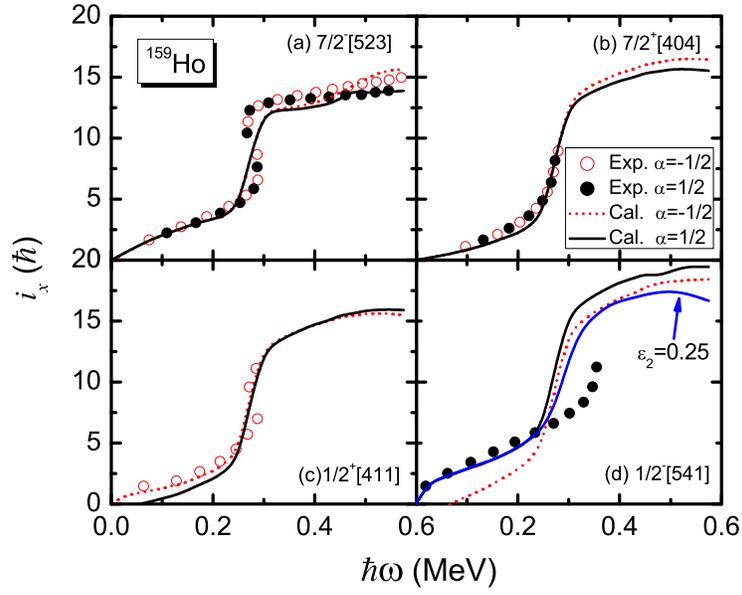}
\caption{\label{fig3:ix} (Color online)
The experimental and calculated alignments of four low-lying
bands in $^{159}$Ho.
The alignments $i_x$ are defined as $i_x= \langle
J_x \rangle -\omega J_0 -\omega ^ 3 J_1$ and the Harris parameters
$J_0 = 23\ \hbar^2$MeV$^{-1}$ and $J_1 = 58\ \hbar^4$MeV$^{-3}$ are
taken from Ref.~\protect\cite{Ollier2011_PRC84-027302}.
The experimental alignments are denoted by black solid circles (signature
$\alpha=+1/2)$ and red open circles (signature $\alpha=-1/2)$, respectively.
The calculated alignments are denoted by black solid lines (signature $\alpha=+1/2)$
and red dotted lines (signature $\alpha=-1/2)$, respectively.
The calculated alignments for  $1/2^-[541]$ ($\alpha =+1/2$) with the deformation
$\varepsilon_2 = 0.25$ are denoted by blue solid line.
}
\end{figure}

Figure~\ref{fig2:MOI}(a)-(d) shows the experimental and calculated
kinematic MOI's $J^{(1)}$ of four low-lying rotational bands in $^{159}$Ho.
The experimental kinematic MOI's, which are extracted from the rotational spectra
in Refs.~\cite{Ma2000_JPG26-43, Ollier2011_PRC84-027302},
are denoted by black solid circles (signature $\alpha=+1/2)$
and red open circles (signature $\alpha=-1/2)$, respectively.
The calculated MOI's by particle-number conserving method are denoted
by black solid lines (signature $\alpha=+1/2)$
and red dotted lines (signature $\alpha=-1/2)$, respectively.
The experimental MOI's of these 1-quasiparticle bands and their variation
with rotational frequency $\hbar\omega$ are well reproduced by the
particle-number conserving calculations except for $1/2^-[541]$,
in which the calculated backbending frequency
($\hbar\omega \sim 0.28$~MeV) is much smaller than the data ($\hbar\omega \sim 0.35$~MeV).
Moreover, the sharp backbending at $\hbar\omega \sim 0.28$~MeV in the experimental MOI's
in Fig.~\ref{fig2:MOI}(a) are not very well reproduced by the calculation.
This is because in the cranking model, before and after the backbending,
the two bands which have very different alignment from each other are mixed.
In order to obtain the backbending effect exactly,
one has to go beyond the cranking model and consider the two quasiparticle configurations
in the vicinity of the critical region~\cite{Hamamoto1976_NPA271-15, Cwiok1978_PLB76-263}.
The experimental and calculated alignments of four low-lying
bands in $^{159}$Ho are also shown in Fig.~\ref{fig3:ix}(a)-(d).
The alignments $i_x$ are defined as $i_x= \langle
J_x \rangle -\omega J_0 -\omega ^ 3 J_1$ and the Harris parameters
$J_0 = 23\ \hbar^2$MeV$^{-1}$ and $J_1 = 58\ \hbar^4$MeV$^{-3}$ are
taken from Ref.~\protect\cite{Ollier2011_PRC84-027302}.
Similar with the MOI's in Fig.~\ref{fig2:MOI}, the experimental alignments
of these 1-quasiparticle bands are also well reproduced by the
particle-number conserving calculations except for $1/2^-[541]$.

It is well known that the first backbending in the odd-$Z$ rare-earth nuclei
is caused by the alignment of one $i_{13/2}$ neutron pair.
As for $^{159}$Ho, in principal, the backbending frequencies ($\hbar\omega \sim 0.28$~MeV)
for these four low-lying bands should be the same.
The delayed crossing frequency in $1/2^-[541]$ is explained by the strongly prolate deformation driving
effect, which has been confirmed from both the theoretical and the experimental
sides~\cite{Nazarewicz1990_NPA512-61, Warburton1995_NPA591-323}.
The calculated results with the deformation $\varepsilon_2 = 0.25$, which is about 10\%
larger than that of the ground state bands of the neighboring even-even nuclei,
are also shown in Fig.~\ref{fig2:MOI}(d) [or Fig.~\ref{fig3:ix}(d)].
It can be seen that, the calculated backbending frequency ($\hbar\omega \sim 0.30$~MeV) is still too small
to reproduce the data ($\hbar\omega \sim 0.35$~MeV) even after increasing the deformation.
The calculated results can only reproduce a delay of about 20~keV,
which is consistent with the cranked Woods-Saxon calculations in Ref.~\cite{Nazarewicz1990_NPA512-61}.
As discussed in Ref.~\cite{Yang1994_CJNP16-217}, except the deformation driving effect,
this delay ($\sim$70~keV) may also caused by other influences, i.e., the residual neutron-proton interaction.

It can be seen from the experimental data in Fig.~\ref{fig2:MOI}(a) [or Fig.~\ref{fig3:ix}(a)]
that not very significant signature splitting
appears in the ground state band $7/2^-[523]$ before the backbending ($\hbar\omega < 0.28$~MeV),
while the splitting is reduced after the first alignment ($\hbar\omega \sim 0.30$~MeV).
At high rotational frequency ($\hbar\omega > 0.40$~MeV) region,
significant signature splitting appears again.
This signature splitting up to the backbending has been explained by
the shape-driving effect of the $h_{11/2}$ quasiprotons~\cite{Frauendorf1983_PLB125-245},
which causes the nucleus to deviate from axial symmetry toward
triaxial shapes with negative $\gamma$ values~\cite{Andersson1976_NPA268-205}.
As discussed in Ref.~\cite{Ollier2011_PRC84-027302},
due to the alignment of $i_{13/2}$ neutrons which drive the nucleus back to axial symmetry,
this splitting is reduced after the first alignment,
and the reemergence of signature splitting at high rotational
frequency region after the backbending
may caused by the negative-$\gamma$ driving influence of the high-$K$ $h_{11/2}$ quasiprotons.
In Ref.~\cite{Hartley1998_PRC58-2720}, the significant signature splitting reemergence
has also been discussed to have a large dependency on
the quadrupole deformation and on the placement of the
proton and neutron Fermi surfaces.
In the particle-number conserving calculations, the significant signature splitting
in the ground state band $7/2^-[523]$ at high rotational frequency
($\hbar\omega > 0.40$~MeV) region
after the backbending is well reproduced,
while there seems no signature splitting before the backbending,
which is inconsistent with the data.
Note that in the present framework of particle-number conserving method,
the triaxial degree of freedom is not considered.
Therefore, the significant signature splitting in the yrast bands of $^{159}$Ho
at high rotational frequency region after the backbending may not come from the
negative-$\gamma$ driving influence of the high-$K$ $h_{11/2}$ quasiprotons
as discussed in Ref.~\cite{Ollier2011_PRC84-027302}.
It can be seen in Fig.~\ref{fig1:Nilsson}(a) that the signature splitting exists
in the cranked single-particle levels of $7/2^-[523]$ at the rotational
frequency $\hbar\omega > 0.40$~MeV and the splitting increases with increasing
rotational frequency.
Therefore, the splitting observed in the yrast
at high rotational frequency region after the backbending
may come from the splitting in the single-particle levels.
Moreover, it is well known that the second backbending in the rare-earth nuclei
is caused by the alignment of one $h_{11/2}$ proton pair.
In the particle-number conserving calculations,
there is no such backbending up to $\hbar\omega \sim 0.55$~MeV,
which is consistent with the data.

It can be seen in Fig.~\ref{fig2:MOI}(b) [or Fig.~\ref{fig3:ix}(b)]
that the particle-number conserving calculations predict a splitting between
the signature partners in $7/2^+[404]$ above $\hbar\omega \sim 0.30$~MeV,
while the experimental data are not available at such high-spin region up to now.
The cranked single-particle levels of $7/2^+[404]$ in Fig.~\ref{fig1:Nilsson}(a)
show that there is no signature splitting in the single particle part.
So it is also interesting to investigate where this splitting comes from.
In the following, the signature splittings in the yrast band $7/2^-[523]$
and the first excited band $7/2^+[404]$ are discussed in detail.

\begin{figure}[h]
\centering
\includegraphics[width=0.7\columnwidth]{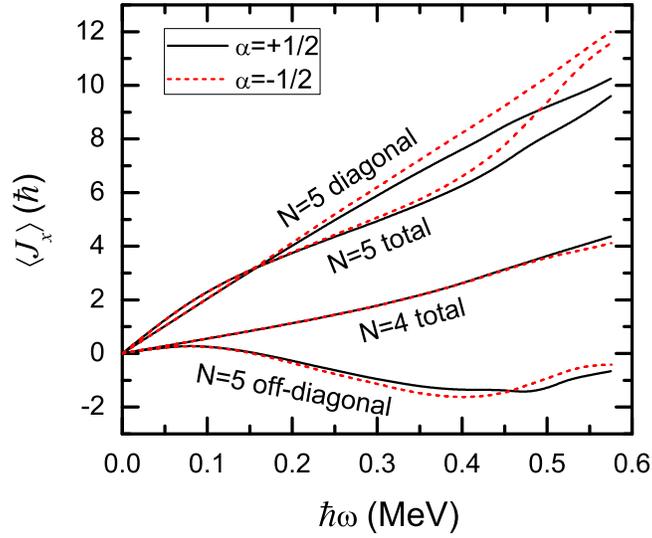}
\caption{\label{fig4:Jx} (Color online)
Contribution of proton $N=4$ and 5 major shell to the angular momentum alignment
$\langle J_x\rangle$ for the ground state band $7/2^-[523]$ in ${}^{159}$Ho.
For the signature $\alpha=+1/2$ band,
the calculated angular momentum alignments are denoted
by black solid lines.
For the signature $\alpha=-1/2$ band,
the calculated angular momentum alignments are denoted red dotted lines.
The contribution of diagonal $\sum_{\mu} j_x(\mu)$ and off-diagonal part
$\sum_{\mu<\nu} j_x(\mu\nu)$ in Eq.~(\protect\ref{eq:jx})
from the proton $N=5$ shell are also shown.
}
\end{figure}

If the proton-neutron residual interaction is neglected,
the splitting between the signature doublets in the odd-$Z$ nuclei
only comes from the contribution of the protons.
In Fig.~\ref{fig4:Jx}, the contribution of proton $N=4$ and 5 major shell
to the angular momentum alignment $\langle J_x\rangle$
for the ground state band $7/2^-[523]$ in ${}^{159}$Ho is shown.
For the signature $\alpha=+1/2$ band,
the calculated angular momentum alignments are denoted
by black solid lines.
For the signature $\alpha=-1/2$ band,
the calculated angular momentum alignments are denoted red dotted lines.
The contribution of diagonal $\sum_{\mu} j_x(\mu)$ and off-diagonal part
$\sum_{\mu<\nu} j_x(\mu\nu)$ in Eq.~(\protect\ref{eq:jx})
from the proton $N=5$ shell are also shown.
Note that in this figure, the smoothly increasing part of the
alignment represented by the Harris formula ($\omega J_0 +\omega ^ 3 J_1$)
is not subtracted.
It can be seen clearly that the signature splitting for
the ground state band $7/2^-[523]$ in ${}^{159}$Ho at $\hbar\omega > 0.40$~MeV
mainly comes from the contribution of the proton $N=5$ shell.
Furthermore, the splitting is mainly from the diagonal part of
the proton $N=5$ shell.

\begin{figure}[h]
\centering
\includegraphics[width=0.7\columnwidth]{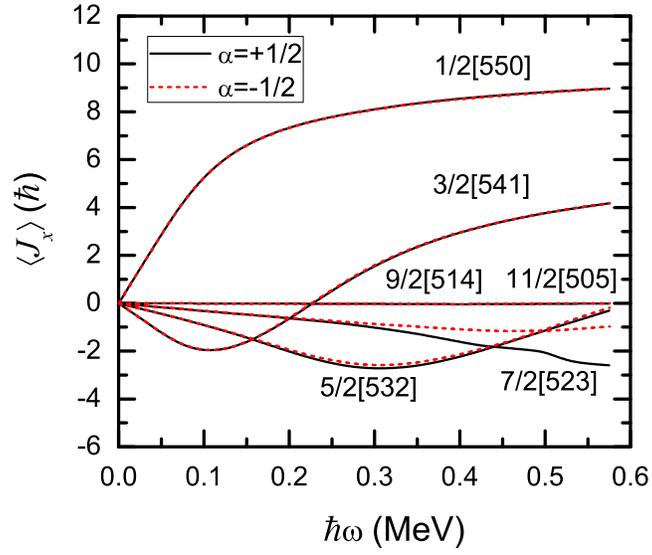}
\caption{\label{fig5:JxOrb} (Color online)
Contribution of each proton orbital form the diagonal part $j_x(\mu)$ in the $N=5$
major shell to the angular momentum alignments
$\langle J_x\rangle$ for the ground state band $7/2^-[523]$ in ${}^{159}$Ho.
For the signature $\alpha=+1/2$ band,
the calculated angular momentum alignments are denoted
by black solid lines.
For the signature $\alpha=-1/2$ band,
the calculated angular momentum alignments are denoted red dotted lines.
}
\end{figure}

In order to have a more clear understanding of the signature splitting in $7/2^-[523]$,
the contribution of each proton orbital from the diagonal part $j_x(\mu)$ in the $N=5$
major shell to the angular momentum alignments $\langle J_x\rangle$ for
the ground state band $7/2^-[523]$ in ${}^{159}$Ho are shown in Fig.~\ref{fig5:JxOrb}.
For the signature $\alpha=+1/2$ band,
the calculated angular momentum alignments are denoted
by black solid lines.
For the signature $\alpha=-1/2$ band,
the calculated angular momentum alignments are denoted red dotted lines.
It can be seen that the contribution of each diagonal part for the
signature doublets are close to each other except for the $7/2^-[523]$,
which shows a splitting from $\hbar\omega \sim 0.30$~MeV, and the splitting
increases with increase rotational frequency.
Therefore, it is clear that the significant splitting
at high rotational frequency ($\hbar\omega > 0.40$~MeV) region
comes from the splitting of the diagonal part of $7/2^-[523]$.

\begin{figure}[h]
\centering
\includegraphics[width=0.7\columnwidth]{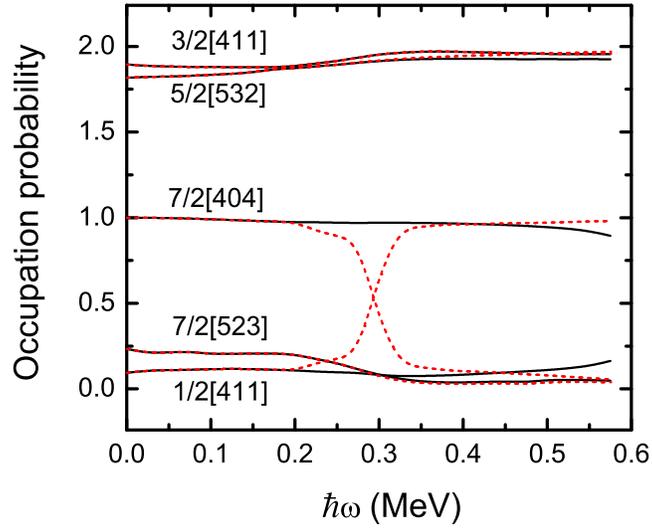}
\caption{\label{fig6:Occup} (Color online)
Occupation probability $n_\mu$ of each orbital $\mu$ near the Fermi surface
for $7/2^+[404]$ with signature $\alpha=+1/2$ (black solid lines)
and $\alpha=-1/2$ (red dotted lines) in $^{159}$Ho.
The Nilsson levels far above the Fermi surface
($n_{\mu}\sim0$) and far below ($n_{\mu}\sim2$) are not shown.
}
\end{figure}

One of the advantages of the particle-number conserving method is that the
total particle number $N = \sum_{\mu}n_\mu$ is exactly conserved,
whereas the occupation probability $n_\mu$ for each orbital varies
with rotational frequency $\hbar\omega$.
By examining the $\omega$-dependence of the orbitals close to the Fermi surface, one
can learn more about how the Nilsson levels evolve with rotation and
get some insights on the level crossings.
In order to learn why there exists a signature splitting in
the first excited band $7/2^+[404]$ at $\hbar\omega > 0.30$~MeV in
the particle-number conserving calculations, the occupation probability $n_\mu$
of each orbital $\mu$ near the Fermi surface
for $7/2^+[404]$ with signature $\alpha=+1/2$ (black solid lines)
and $\alpha=-1/2$ (red dotted lines) in $^{159}$Ho are shown in Fig.~\ref{fig6:Occup}.
The Nilsson levels far above the Fermi surface
($n_{\mu}\sim0$) and far below ($n_{\mu}\sim2$) are not shown.
It can be seen that, for the signature $\alpha=+1/2$ band, the
$7/2^+[404]$ always occupied from the beginning to the end,
while for the signature $\alpha=-1/2$ band, the occupation
probability drops drastically to nearly zero at $\hbar\omega \sim 0.30$~MeV
and the $1/2^+[411]$ orbital becomes occupied ($n_\mu \sim 1$).
This indicates that there exists a level crossing between $7/2^+[404]$ ($\alpha=-1/2$)
and $1/2^+[411]$ ($\alpha=-1/2$) at $\hbar\omega \sim 0.30$~MeV.
If we look at the cranked Nilsson levels in Fig.~\ref{fig1:Nilsson}(a), we can see that
the cranked single-particle level of $1/2^+[411]$ ($\alpha=-1/2$) drops with
increasing rotational frequency and acrosses $7/2^+[404]$ at $\hbar\omega \sim 0.20$~MeV.
Note that in the cranked single-particle levels, the pairing interaction is not included.
After taking into account of the pairing interaction, the level crossing frequency
will be delayed. Due to the consistency of the calculated bandhead energies of
these 1-quasiparticle states with the data,
this level crossing in the particle-number conserving calculation is reasonable.
Therefore, the splitting in $7/2^+[404]$ above $\hbar\omega \sim 0.30$~MeV may
come from the level crossing with $1/2^+[411]$.

\begin{figure}[h]
\centering
\includegraphics[width=0.7\columnwidth]{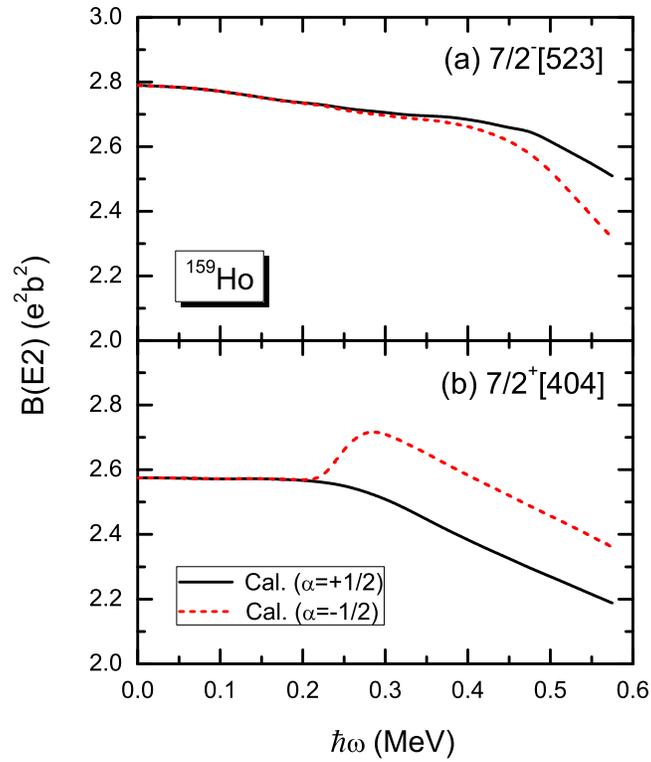}
\caption{\label{fig7:BE2} (Color online)
The calculated $B(E2)$ values for (a) $7/2^-[523]$ and (b) $7/2^+[404]$.
The calculated $B(E2)$ values are denoted by black solid lines (signature $\alpha=+1/2)$
and red dotted lines (signature $\alpha=-1/2)$, respectively.
}
\end{figure}

The $B(E2)$ transition probabilities are important quantities
to test the structure changes in a rotating nucleus,
i.e., the splittings in $7/2^-[523]$ and $7/2^+[404]$,
and can give a crucial test for the present model.
Especially for $7/2^+[404]$, crossing of two bands with different configurations
can lead to a structure change before and after the crossing,
which can give rise to observable effects.
Using the semiclassical approximation, the $B(E2)$
transition probabilities can be obtained according to Eq.~(\ref{eq:be2}).
Note that in the framework of particle-number conserving method,
the $B(E2)$ transitions of the antimagnetic rotation bands in
$^{105, 106}$Cd have already been investigated and
the data are reproduced quite well~\cite{Zhang2013_PRC87-054314}.
The calculated $B(E2)$ values for $7/2^-[523]$ and $7/2^+[404]$ are shown in Fig.~\ref{fig7:BE2}.
The calculated $B(E2)$ values are denoted by black solid lines (signature $\alpha=+1/2)$
and red dotted lines (signature $\alpha=-1/2)$, respectively.
It can be seen in Fig.~\ref{fig7:BE2}(a) that, the $B(E2)$ values
for the signature doublets in $7/2^-[523]$ are similar with each other at low
rotational frequency, while the signature $\alpha=-1/2$ band drops more quickly
than the signature $\alpha=+1/2$ band at $\hbar\omega > 0.30$~MeV.
It is understandable that the difference in $B(E2)$ values just comes from
the difference in the wave-functions in this signature doublets, which becomes
larger at higher rotational frequency.
In Fig.~\ref{fig7:BE2}(b), the $B(E2)$ values the $7/2^+[404]$ with signature $\alpha=-1/2$
jump at $\hbar\omega \sim 0.30$~MeV and then decrease with
increasing rotational frequency.
Obviously, this jump in the signature $\alpha=-1/2$ band comes from
the level crossing with $1/2^+[411]$, which lead to a structure change in
the many-body wave functions.
Note that the valence single-particle space is constructed
in the major shells from $N=0$ to $N=5$ for protons
when the $B(E2)$ values are calculated,
so there is no effective charge involved in the calculation of the $B(E2)$ values.
Due to experimental difficulties, little information on transition properties can be
found for ${}^{159}$Ho.
Therefore, these calculations on $B(E2)$ transition probabilities
may be suggested for future experiments.

\section{Summary}
\label{sec:summ}

The high-spin rotational bands in odd-$Z$ nuclei $^{159}$Ho are investigated
by using the cranked shell model with the pairing correlations treated
by a particle-number conserving method,
in which the blocking effects are taken into account exactly.
The experimental moments of inertia and alignments
and their variations with the rotational frequency $\hbar\omega$
are reproduced very well by the present calculations.
The signature splitting between the signature partners in the
yrast band $7/2^-[523]$ is discussed and the splitting in $7/2^+[404]$
above $\hbar\omega \sim 0.30$~MeV is predicted due to the level
crossing with $1/2^+[411]$.
To test the predictions, the calculated $B(E2)$ transition probabilities
are suggested for future experiments.

\section{Acknowledgement}
\label{acknowledgement}

Helpful discussions with En-Guang Zhao and Shan-Gui Zhou are gratefully acknowledged.
This work was partly supported by the Fundamental Research Funds for the Central Universities (2015QN21),
and the National Natural Science Foundation of China (Grants No. 11275098, 11275248, 11505058).
The computational results presented in this work have been obtained on the High-performance
Computing Cluster of SKLTP/ITP-CAS and the ScGrid of the Supercomputing Center,
Computer Network Information Center of the Chinese Academy of Sciences.






\newpage
\begin{table}[!h]
\tabcolsep 0pt
\caption{\label{tab:lev} The experimental and the calculated
bandhead energies of 1-quasiparticle bands in $^{159}$Ho.
The calculated results using Nilsson parameters in Ref.~\cite{Nilsson1969_NPA131-1}
and the modified values  are denoted by $E_{\rm Cal}^{\rm Lund}$ and
$E_{\rm Cal}^{\rm New}$, respectively. }
\vspace*{-12pt}
\begin{center}
\def\temptablewidth{0.8\textwidth}
{\rule{\temptablewidth}{1pt}}
\begin{tabular*}{\temptablewidth}{@{\extracolsep{\fill}}cccc}
 Configuration & $E_{\rm Exp}$ (keV) &$E_{\rm Cal}^{\rm Lund}$ (keV) & $E_{\rm Cal}^{\rm New}$ (keV) \\
 \hline
 $7/2^-[523]$  & 0                   &  0                            &  0      \\
 $7/2^+[404]$  & 166                 &  345                          &  174    \\
 $1/2^+[411]$  & 206                 &  401                          &  211    \\
 $5/2^+[402]$  & 253                 &  801                          &  612
\end{tabular*}
{\rule{\temptablewidth}{1pt}}
\end{center}
\end{table}

\newpage

\newpage

\newpage

\newpage

\newpage

\newpage

\newpage

\end{document}